\documentclass{jetpl}
\twocolumn
\lat

\usepackage{graphicx}
\usepackage{epstopdf}
\usepackage{color}
\usepackage{srctex}
\usepackage{hyperref}
\usepackage{amsmath,amssymb}

\usepackage{widetext}
\usepackage{flushend}
\usepackage{cuted}

\DeclareMathOperator{\Tr}{Tr}
\DeclareMathOperator{\RS}{\mathrm{\scriptscriptstyle RS}}\DeclareMathOperator{\1RSB}{\mathrm{\scriptscriptstyle 1RSB}}
\DeclareMathOperator{\RSB}{\mathrm{\scriptscriptstyle 1RSB}}

\title{Full replica symmetry breaking in p-spin-glass-like  systems }

\rtitle{FRSB in p-spin-glass-like systems\ldots}

\sodtitle{FRSB in p-spin-glass-like systems}

\author{T.~I.~Schelkacheva$^{1}$, and N.~M.~Chtchelkatchev$^{1-6}$}
\rauthor{T.~I.~Schelkacheva, and N.~M.~Chtchelkatchev}

\sodauthor{Schelkacheva, Chtchelkatchev}
\address{$^1$Institute for High Pressure Physics, Russian Academy of Sciences, 142190, Troitsk, Moscow, Russia
\\~\\
$^2$ L.D. Landau Institute for Theoretical Physics, Russian Academy of Sciences, 142432, Moscow Region, Chernogolovka, Russia
\\~\\
$^3$Ural Federal University, 620002 Ekaterinburg, Russia
\\~\\
$^4$Department of Theoretical Physics, Moscow Institute of Physics and Technology, 141700 Moscow, Russia
\\~\\
$^5$All-Russia Research Institute of Automatics, 22 Suschevskaya, Moscow 127055, Russia
\\~\\
$^6$Institute of Metallurgy, Ural Division of Russian Academy of Sciences, Yekaterinburg 620016, Russia
}

\abstract{

It is shown that continuously changing the effective number of interacting particles in p-spin-glass-like model allows to describe the transition from the full replica symmetry breaking glass solution to stable first replica symmetry breaking glass solution in the case of  non-reflective symmetry diagonal operators used instead of Ising spins. As an example, axial quadrupole moments in place of Ising spins are considered and the boundary value $p_{c_{1}}\cong 2.5$  is found.
}

\begin{document}

\maketitle
\paragraph*{Introduction\label{Sec:Intro}}

 The basis of understanding glasses  is the Sherrington -- Kirkpatrick (SK) model~\cite{sk}: the
Ising model with random links. A stable solution for SK model  was obtained by Parisi~\cite{par1,book} with a full replica symmetry breaking (FRSB) scheme. Later it was realised that replica symmetry is not abstract and academic question but it corresponds to formation of the specific hierarchy of basins in the energy landscape of the glass forming system.

A natural generalization of the SK model with pair interaction of spins is a model with $p$-spin interactions~\cite{book,Gardner}. Unlike of the SK model, $p$-spin model has a stable first replica symmetry breaking (RSB) solution that arises abruptly. A very low-temperature boundary of the 1RSB stability region is given by so called Gardner transition temperature intensively discussed last time~\cite{Gardner,F.Zamponi} where a valley in configuration space transforms to a multitude of separated basins.

Now there is a reborn of interest to spin models showing glassy behaviour~\cite{F.Zamponi,Cris,Kirk,Kirkp,Franz,F.Caltagirone,Riz,Rizzo,Par,B}. It turned out that these models can qualitatively explain physics of ``real'' glasses~\cite{Wolynes}. On the other hand, there is limited number of analytically solvable glassy models and each such model is interesting itself. Here we propose analytical solution of p-spin-glass-like  system and discuss physical applications of this model.


For a long time there was a conjecture that there are more or less two classes of models, depending on how the replica symmetry breaking appears~\cite{PRE}. In one class of models full replica symmetry breaking (FRSB) occurs continuously at the transition point from the paramagnetic state to the glass state (like, e.g., in SK model). The second class of models can be called 1RSB-models ($p$-spin model, Potts models). In this case there is a finite range of temperatures where stable 1RSB glass solution occurs. What is important that this 1RSB solution mostly appears abruptly.

1RSB-models and especially $p$-spin glasses in recent years attract much interest in connection with the fact that there is a close relationship between static replica approach and dynamic consideration. For example the Random First Order Transition theory for structural glasses is inspired by the $p$-spin glass model~\cite{F.Zamponi,Cris,Kirk,Kirkp,Franz,F.Caltagirone,Riz,Rizzo,Par,B,Wolynes}.

It should be also noted that the two classes of models mentioned above distinguish essentially by their energy landscape~\cite{U.Buchenau}, which is an important concept in the dynamics of liquids and glasses.
\begin{figure}[b]
  \centering
  \includegraphics[width=0.9\columnwidth]{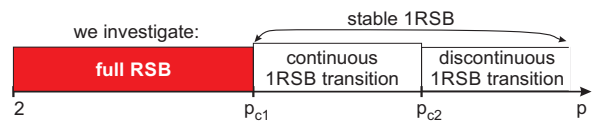}\\
  \caption{Fig.1 We have got FRSB glass solution of a p-spin-like model when the effective number of interacting particles $2<p<p_{c1}$. We considered general diagonal operators $\hat{U}$ with $\Tr\hat{U}^{2k+1} \neq 0$, $k=1,2,...$ instead of Ising spins.\label{fig0}}
\end{figure}

In the context of SK-like and 1RSB-models it is possible  to develop very advanced theoretical tools that can be reused in other contexts. These relatively simple models based on well-designed solutions allow to explore qualitatively an extensive range of issues, ranging from magnetic to structural glass transitions~\cite{F.Zamponi,Cris,Kirk,Kirkp,Franz,F.Caltagirone,Riz,Rizzo,Par,B,Wolynes,U.Buchenau,Alba}.

Therefore, a natural generalizations of these basic models leads to a successful description of the various types of glasses, such as orientation glasses and cluster glasses.  Replacing the simple Ising spins by more complex operators $\hat{U}$ dramatically expands the range of solvable tasks. The operator $\hat{U}$ can be, for example, the axial quadrupole moment (quadrupole  glass in molecular solid hydrogen at different pressures)  or the role of $\hat{U}$ can be played by certain combinations of the cubic harmonics (orientational glass of the clusters, for example, in $C_{60}$ in a wide pressure range), see~\cite{PRE,Prb,C60,L,Chtch,W,ESc,EShc,JPA}.

It was shown recently that many statements related to the models based on Ising spins can be applied to $p$-spin-like models that use instead of spins diagonal operators $\hat{U}$ when there is \textit{``reflection symmetry''}: $\Tr\hat{U}^{2k+1} =0$, $k=1,2...$. For such models we can build a stable 1RSB solution for $p>2$ in a wide region of temperatures. It was shown that the point $p = 2$ is special  for such models~\cite{JPA,schelk,Full}.

For p-spin-like model we previously have got significantly different solutions when diagonal operators $\hat{U}$ have \textit{broken reflective symmetry}, then $\Tr\hat{U}^{2k+1} \neq 0$ like (e.g., for quadrupole operators). And for these models 1RSB glass solution behavior have been recently investigated near the glass transition at different (continuous) $p$~\cite{JPA,EShch}. It turned out that there is a finite region of instability of 1RSB solutions for $2\leq p<p_{c_{1}}$ where $p_{c_{1}}$ is determined by the specific form of $\hat{U}$. 1RSB solution is stable for $p>p_{c_{1}}$. Wherein the transition from para-phase to 1RSB glass  is continuous for $p_{c_{1}}< p<p_{c_{2}}$. When $p>p_{c_{2}}$~\cite{JPA,EShch} 1RSB glass occurs abruptly just as in the conventional $p$-spin model. We should note that $p_{c_{2}}$ is not universal, but it depends on the particular  type of $\hat{U}$. We have built~\cite{sche,schel} FBSB solution for these models with a pair interaction $p=2$.

In this letter  we  investigate in detail the generalised $p$-spin glass forming models in the region $2\leq p<p_{c_{1}}$ where  instability of 1RSB glass solution is expected. We build a solution with full replica symmetry breaking.  The very existence of the domain with full replica symmetry breaking is a surprising result especially compared with the traditional $p$-spin model of Ising spins. Continuously changing the effective number of interacting particles in $p$-spin model allows us to describe the crossover from the full replica symmetry breaking glass solution to stable first replica symmetry breaking glass solution (in our case of  non-reflective symmetry diagonal operators used instead of Ising spins).  For illustrating example we take operators of axial quadrupole moments in place of Ising spins. For this model  we find the boundary value $p_{c_{1}}\cong 2.5$.

\paragraph*{The model\label{secmodel}}

The staring point is the p-spin-glass-like  Hamiltonian
\begin{equation}
H=-\sum_{{i_{1}}\leq{i_{2}}...\leq{i_{p}}}J_{i_{1}...i_{p}}
\hat{U}_{i_{1}}\hat{U}_{i_{2}}...\hat{U}_{i_{p}}, \label{one}
\end{equation}
where the quenched interactions $J_{i_{1}...i_{p}}$ are distributed with Gaussian probability:
\begin{equation}
P(J_{i_{1}...i_{p}})=\frac{\sqrt{N^{p-1}}}{\sqrt{p!\pi}
J}\exp\left[-\frac{(J_{i_{1}...i_{p}})^{2}N^{p-1}}{ p!J^{2}}\right]. \label{two}
\end{equation}
Here arbitrary non-reflective symmetry diagonal operators $\hat{U}$ are located on  the lattice sites $i$ instead of Ising spins, $N$ is the number of sites. We remind that it implies that $\Tr \hat U^{2k+1}\neq 0$, k$=1,2,\ldots$. Here we investigate how order parameters of this model develop with the continuous parameter $p$ and the specific type of the operators
$\hat{U}$.

We write down in standard way the disorder averaged free energy using the replica approach~\cite{book}.
We assume like it was done in~\cite{sche, schel} that the order parameter deviations $\delta q^{\alpha \beta}$ are small from replica symmetry order parameter $q_{\RS}$. To describe the RSB-solution near the  bifurcation point temperature $T_0$, where RS ``transforms'' into RSB, we expand the expression for the free energy up to the fourth order of $\delta q^{\alpha \beta}$. We write below in~\eqref{10frs} the deviation $\Delta F(p)$ of the free energy  from its RS-part. It is important to note that the expression for Free energy have not been written before for arbitrary $p$. We should also emphasize that it includes the terms with odd number of identical replica indices~\cite{Tem, T.Temesvari, C.DeDominicis} unlike Ising-spin SK-models.

All coefficients in~\eqref{10frs} depend only on RS-solution at $T_{0}$. They are given in Supplementary Material. The prime on $\sum'$ means that only the superscripts belonging to the same  $\delta q$ are necessarily different. The coefficients of the second and third orders, $\lambda_{\rm(\RS)\,repl}, L, B_{3},..B_{4}$, have been obtained earlier~\cite{PRE, JPA,schelk}, but we also write
them in Supplementary Material for readability. This is the only overlap with our previous publications.

\begin{widetext}
\begin{multline}\label{10frs}
\frac{\Delta F(p)}{NkT}=\lim_{n \rightarrow 0}\frac{1}{n}\Biggl\{\frac{t^2}{4}\frac{p(p-1)}{2}{q^{(p-2)}_{\RS}}\left[\lambda(p)_{\rm(\RS)\, repl}\right]
{\sum_{\alpha,\beta}}^{'}\left(\delta q^{\alpha\beta}\right)^{2}-\frac{t^{4}}{2}L(p){\sum_{\alpha,\beta,\delta}}^{'}\delta q^{\alpha\beta}\delta
q^{\alpha\delta}- t^{6}\biggl[B_{2}(p){\sum_{\alpha,\beta,\gamma,\delta}}^{'}\delta q^{\alpha\beta}\delta q^{\alpha\gamma}\delta q^{\beta\delta}+
\\
B'_{2}(p){\sum_{\alpha,\beta,\gamma,\delta}}^{'}\delta q^{\alpha\beta}\delta q^{\alpha\gamma}\delta q^{\alpha\delta}+
B_{3}(p){\sum_{\alpha,\beta,\gamma}}^{'}\delta q^{\alpha\beta}\delta q^{\beta\gamma}\delta q^{\gamma\alpha}+B'_{3}(p){\sum_{\alpha,\beta,\gamma}}^{'}\left(\delta
q^{\alpha\beta}\right)^{2}\delta q^{\alpha\gamma}+B_{4}(p){\sum_{\alpha,\beta}}^{'}\left(\delta q^{\alpha\beta}\right)^{3}\biggr]+
\\
t^{8}\biggl[D_{2}(p){\sum_{\alpha,\beta}}^{'}\left(\delta q^{\alpha\beta}\right)^{4}+D_{31}(p){\sum_{\alpha,\beta,\gamma}}^{'}\left(\delta q^{\alpha\beta}\right)^{3}\delta q^{\alpha\gamma} + D_{32}(p){\sum_{\alpha,\beta,\delta}}^{'}(\delta q^{\alpha\beta})^{2}\left(\delta q^{\alpha\delta}\right)^{2}+D_{33}(p){\sum_{\alpha,\beta,\gamma}}^{'}\left(\delta
q^{\alpha\beta}\right)^{2}\delta q^{\alpha\gamma}\delta q^{\gamma\beta}+
\\
D_{42}(p){\sum_{\alpha,\beta,\gamma,\delta}}^{'}\left(\delta q^{\alpha\beta}\right)^{2}\delta q^{\alpha\gamma}\delta q^{\alpha\delta}+D_{43}(p){\sum_{\alpha,\beta,\gamma,\delta}}^{'}
\left(\delta q^{\alpha\beta}\right)^{2}\delta q^{\alpha\gamma}\delta q^{\beta\delta}+
D_{45}(p){\sum_{\alpha,\beta,\gamma,\delta}}^{'}\left(\delta q^{\alpha\beta}\right)^{2}\delta q^{\alpha\gamma}\delta q^{\gamma\delta}+
\\
D_{46}(p){\sum_{\alpha,\beta,\gamma,\delta}}^{'}\delta q^{\alpha\beta}\delta q^{\alpha\gamma}\delta q^{\alpha\delta}\delta q^{\beta\gamma}+
D_{47}(p){\sum_{\alpha,\beta,\gamma,\delta}}^{'}\delta q^{\alpha\beta}\delta q^{\beta\gamma}\delta q^{\gamma\delta}\delta q^{\delta\alpha}+D_{53}(p){\sum_{\alpha,\beta,\gamma,\delta,\mu}}^{'}\delta q^{\alpha\beta}\delta q^{\alpha\gamma}\delta q^{\alpha\delta}q^{\alpha\mu}+
\\
D_{54}(p){\sum_{\alpha,\beta,\gamma,\delta,\mu}}^{'}\delta q^{\alpha\beta}\delta q^{\alpha\gamma}\delta q^{\alpha\delta}q^{\beta\mu}+
D_{55}(p){\sum_{\alpha,\beta,\gamma,\delta,\mu}}^{'}\delta
q^{\alpha\beta}\delta q^{\alpha\gamma}\delta q^{\gamma\delta}q^{\delta\mu}\biggr]\Biggr\},
\end{multline}
\end{widetext}

The order parameters and the replicon mode $\lambda$ (responsible for RSB stability) we find as follows:
\begin{multline}\label{lambdaRS}
\lambda(p)_{\rm (\RS)\, repl}=1 - t^{2}\frac{p(p-1)}{2}{q^{(p-2)}_{\RS}}\times
\\
\int dz^G \left\{\frac{\Tr\left(\hat{U}^2 \exp{\hat{\theta}_{\RS}}\right)} {\Tr \exp{\hat{\theta}_{\RS}}}- \left[\frac{\Tr\hat{U} \exp{\hat{\theta}_{\RS}}}
{\Tr \exp{\hat{\theta}_{\RS}}}\right]^2\right\}^2;
\end{multline}
\begin{gather}\label{0qrs}
q(p)_{\RS}=\int dz^G\left\{ \frac{\Tr\left[\hat{U} \exp\left(\hat{\theta}_{\RS}\right)\right]}
{\Tr\left[\exp\left(\hat{\theta}_{\RS}\right)\right]}\right\}^{2};
\\
w(p)_{\RS}=\int dz^G \frac{\Tr\left[{\hat{U}}^2 \exp\left(\hat{\theta}_{\RS}\right)\right]}
{\Tr\left[\exp\left(\hat{\theta}_{\RS}\right)\right]};
\end{gather}
where  $\int dz^G = \int_{-\infty}^{\infty} \frac{dz}{\sqrt{2\pi}}\exp\left(-\frac{z^2}{2}\right)$, $t={J}/kT=t_{0}+\Delta t$; $\alpha, \beta$ label $n$ replicas and
\begin{gather}
\notag
\hat{\theta}(p)_{\RS}=zt\sqrt{\frac{p{q_{\RS}}^{(p-1)}}{2}}\,\hat{U}+t^2\frac{p[{w_{\RS}}^{(p-1)}-
{q_{\RS}}^{(p-1)}]}{4}\hat{U}^2.
\end{gather}

\subparagraph*{FRSB}
Below we find answer of two questions: 1) if FRSB realises for given model (we derive simple criterium), 2) we provide necessary tools for calculation of FRSB order parameter $q(x)$.

To describe the FRSB function $q(x)$  of the variable $x$ we include in the consideration the fourth-order terms in the expansion of $\Delta F$. We use the standard formalized Parisi algebra rules~\cite{par1,book} to write the free energy as the functional of $q (x)$ and so to construct FRSB. The terms which are not formally described by the Parisi rules can be reduced to the standard form as well~\cite{sche, schel}.

The equation for the order parameter $q(x)$ follows from the  stationarity condition $\frac\delta{\delta q(x)}\Delta F(p) = 0$.

Therefore, we can is a similar way find  the branching condition (appearance of FRSB glass solution), which were derived in detail for $p=2$ (see \cite{sche, schel} and Refs. therein):
$\lambda(p)_{\rm(\RS)\, repl}\mid_{T_{0}}=0$,  which produces the temperature $T_{0}(p)$.

For clarity we write $\frac\delta{\delta q(x)}\Delta F(p) = 0$ up to the terms of the second order. We get
\begin{multline}\label{1340frs}
-\frac{t_{0}^2}{2}\frac{p(p-1)}{2}{q^{(p-2)}_{\RS}}\frac{d\left[\lambda_{\rm(\RS)\, repl}\right]}{dt}|_{t_{0}}\Delta t
q(x)-
\\
t_{0}^{4}L\langle q\rangle +...=0.
\end{multline}

So if the operators $\hat{U}$ do not have the reflective symmetry (therefore $L\neq 0$ in that case) we need an additional branching condition,
$\langle q\rangle\equiv\int_{0}^{1}q(x)dx =0+o(\Delta t)^{2}$~\cite{schel}, insuring the appearance of non-trivial new solutions.

Integral equation, $\frac\delta{\delta q(x)}\Delta F = 0$, that  determines the function $q(x)$, as  usually can be simplified
using the differential operator $\hat{O}=\frac{1}{q'}\frac{d}{dx}\frac{1}{q'}\frac{d}{dx}$, where
 $q'= \frac{d q(x)}{dx}$:
\begin{multline}\label{ten7688}
t^{6}\left\{B_{4}-B_{3}x\right\}+t^{8}\Biggl\{\left[-2D_{33}+4xD_{47}\right]\biggl[-xq(x)-
\\
\int_{x}^{1}dy q(y) \biggr]+
\left[-4D_{2}+2xD_{33}\right]q(x)\Biggr\}=0\,.
\end{multline}
The coefficients are given in Supplementary Material for $p\geq 2$.

Since $q(x)$ can only be a non-decreasing function of $x$ we should consider how the sign of $q'= \frac{d q(x)}{dx}$ depends on the parameter $p$. We obtain from Eq.~(\ref{ten7688}):
\begin{equation}\label{ten73}
q'(p)=\frac{B_{3}+4[xq(x)-\int_{0}^{x}dy q(y)]D_{47}+2D_{33}q(x)}{t^{2}4\left[-D_{2}+D_{33}x-D_{47}x^{2}\right]}.
\end{equation}
This expression is one of the central results of our paper. Depending on the sign of $q'$ we can conclude if the system falls into FRSB (positive $q'$) state or not.

If we confine ourselves to the terms of the third order over $\delta q$ in $\Delta F$ and \eqref{ten7688} then we obtain $t^{6}\left\{B_{4}-B_{3}x\right\}+...=0$.
So the significantly depending on $x$ part of $q(x)$ is concentrated in the neighborhood of
\begin{gather}\label{ten9888}
\tilde{x}(p)={{B_{4}}(p)/{B_{3}}}(p).
\end{gather}
Only in the case of operators $\hat{U}$ with $\Tr{\hat{U}}^{(2k+1)}\neq0$ for $k=1,2,..$ we get $\tilde{x}={{B_{4}}/{B_{3}}}\neq 0$. For Ising-like operators with $\Tr{\hat{U}}^{(2k+1)}=0$ we get $p=2$ (see ~\cite{ JPA,schelk, EShch}), $B_{4}=0$ and so $\tilde{x}=0$ in accordance with usual standard case of the Parisi theory~\cite{par1,book}.

\paragraph*{Example\label{secFRSB}}
\begin{figure}[tb]
  \centering
  \includegraphics[width=0.99\columnwidth]{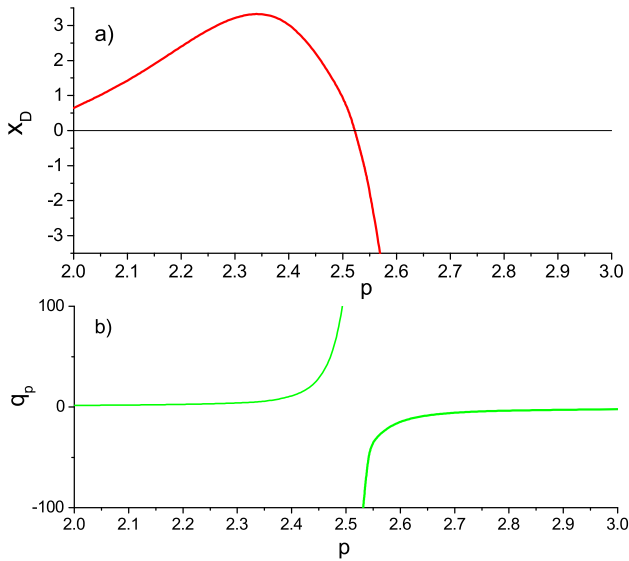}\\
  \caption{Fig.~2 a).The dependence of the denominator  $x_D=\left[-D_{2}+D_{33}x-D_{47}x^{2}\right]\mid_{\tilde{x}}$ of the function $\frac{dq(x)}{dx }\mid_{\tilde{x}}$ on the effective number of interacting quadrupoles $p$ obtained from Eq.~(\ref{ten73}). b) The dependence of the function $q_p=\frac{dq(x)}{dx }\mid_{\tilde{x}}$ on the effective number of interacting quadrupoles $p$ obtained from Eq.~(\ref{ten73}).}\label{fig1}
\end{figure}

As an illustrating example, we  consider quadrupole glass in the space, $J = 1$. Operator $\hat U =\hat Q$ is the axial quadrupole moment and it takes values $(-2, 1, 1)$. So there is no reflection symmetry. Behavior of this model is significantly different from what one usually has in the case of Ising-like operators~\cite{par1,book,PRE,Prb,C60,L,Chtch,W,ESc,EShc,JPA,Full}.

We have done earlier calculations of 1RSB solution for various values of $p$ that could change continuously~\cite{JPA}. It has been found that for $2 <p <p_{c_{1}}$ ($p_{c_{1}}= 2.5$) the solution is qualitatively similar  to one which occurs in the case of pair interaction ($p=2$). This behavior is natural for continuity reason. At high temperatures $T>T_{0}$ (we remind that $T_0$ is the  bifurcation temperature) there is a stable non-trivial solution for the RS order parameter $q_{\RS}$. But $q_{\RS}\neq 0$ for any finite temperature $T>T_{0}$ unlike conventional Ising-spin  model. For $T<T_{0}$ we have Almeida-Thoulles replicon mode $\lambda_{\rm(\RS)\, repl}<0$ and RS-solution becomes unstable. Similarly, when $T<T_{0}$ we have $\lambda_{\rm(\1RSB)\, repl}<0$ determining the condition when the 1RSB glass is unstable~\cite{Prb,ESc,JPA,schelk}.

\begin{figure}[tb]
  \centering
  \includegraphics[width=0.9\columnwidth]{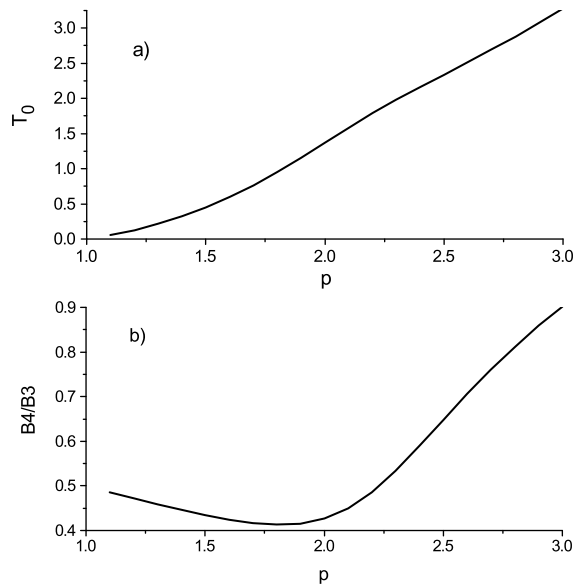}\\
  \caption{Fig.~3 a) Dependence of branching temperature $T_{0}$ of glass solution on the effective number of interacting quadrupoles $p$. b) Dependence of $\tilde{x}(p)={B_{4}(p)}/{B_{3}(p)}$ on the effective number of interacting quadrupoles $p$.}\label{fig2}
\end{figure}
So we investigated the stability of 1RSB solution and found that 1RSB solution is unstable when $p<2.5$. The 1RSB solution becomes stable when $p> 2.5$, see Fig.~\ref{fig0}. The instability region of 1RSB solution is an area in which there may be FRSB solution.

Using Eqs.~(\ref{ten73})-(\ref{ten9888}), we can describe FRSB solution for $T<T_{0}$ near $T_{0}$ for $2 <p <p_{c_{1}}$. We also want to find the value of $p_{c_{1}}$ just numerically solving the equations Eqs.~(\ref{ten73})-(\ref{ten9888}). We intend to compare this result of $p_{c_{1}}$  to $p_{c_{1}}=2.5$ we had previously received, considering the stability of the 1RSB solution~\cite{JPA}.

For pair interaction, $p=2$, we earlier obtained~\cite{schel} that $T_{0} = 1.37$ and $\tilde{x} = 0.43$ for FRSB. Wherein $q=q(x)$ is a continuous increasing function of $x$ in narrow neighborhood, e.g., $\Delta x=0.016$ near $\tilde{x}$ for $(T-T_{0}) = -0.2$. For other values of $x$, not in the neighborhood of $\tilde{x}$, $q(x)$ is very close to $q_{\RSB}$.

So in our case $2 <p <\cong 2.5$ we can with good accuracy put $q(\tilde{{x}})=q(\tilde{{x}})_{\RSB}= q({0})_{\RSB}$ and $[\tilde{x}q(\tilde{x})-\int_{0}^{\tilde{x}}dy q(y)]= 0$ in \eqref{ten73}.  We also have got that $q(\tilde{x})_{\RSB}$ is a small quantity and so we can neglect $2D_{33}q({\tilde{x}})$ compared with $B_{3}$. For other values of $x$ not very close to $\tilde{x}$ in exactly the same way as it was obtained for $p=2$~\cite{schel} we find that $q(x)$ is very close to $q_{\RSB}$.

The results of our calculations are presented in Figs.~\ref{fig1}-\ref{fig2}. At $p_{c_{1}}\cong 2.5$ the function $\frac{dq(x)}{dx }\mid _{\tilde{x}}$ diverges and changes its sign when $p>p_{c_{1}}$ contrary to its conventional probabilistic interpretation. It follows that FRSB is impossible for $p>\approx 2.5$. This result agrees with $p_{c_{1}}$ obtained by us previously for 1RSB solution stability border~\cite{JPA}.

For clarity we show the dependence the denominator   $x_D=\left[-D_{2}+D_{33}x-D_{47}x^{2}\right]\mid_{\tilde{x}}$ on $p$ in Fig.~\ref{fig1}. We show in Fig.~\ref{fig2} the dependence of the branching temperature $T_{0}$  and the dependence of $\tilde{x}(p)={{B_{4}}(p)/{B_{3}}}(p)$ on the effective number of interacting quadrupoles $p$.

\paragraph*{Conclusions \label{Sec:Conc}}

For the first time it is shown that continuously changing of one of the parameters of glass forming model with one type of interaction allows moving from FRSB glass to stable 1RSB solution. As an example the generalized p-spin quadrupole glass model is considered.

\paragraph*{Acknowledgments}
This work was supported in part by the Russian Foundation for Basic Research ( No. 16-02-00295 ) while numerical simulations were funded by Russian Scientific Foundation (grant No. 14-12-01185).


\end{document}